\documentclass[10pt,conference]{IEEEtran}
\usepackage{graphicx}
\usepackage[normalem]{ulem}
\usepackage{amsmath}
\usepackage{bbm}
\usepackage[caption=false]{subfig}
\usepackage{tikz}
\usetikzlibrary{quantikz2}
\usepackage{float}
\usepackage{algorithm}
\usepackage{algpseudocode}

\usepackage[colorlinks=true,urlcolor=blue, citecolor=blue, linkcolor=blue, plainpages=false]{hyperref}  

\title{Algorithmic Advances Towards a Realizable Quantum Lattice Boltzmann Method}
\author{\IEEEauthorblockN{Apurva Tiwari\IEEEauthorrefmark{1}\textsuperscript{\textsection},
Jason Iaconis\IEEEauthorrefmark{2}\textsuperscript{\textsection},
Jezer Jojo\IEEEauthorrefmark{1}\textsuperscript{\textsection},
Sayonee Ray\IEEEauthorrefmark{2}\textsuperscript{\textsection},
Martin Roetteler\IEEEauthorrefmark{2},
Chris Hill\IEEEauthorrefmark{1},
Jay Pathak\IEEEauthorrefmark{1}}
\IEEEauthorrefmark{1}\textit{Ansys Inc., USA.} \\
\IEEEauthorrefmark{2}\textit{ IonQ Inc., 4505 Campus Dr., College Park, MD 20740, USA.}}
\date{}

\begin{document}

\maketitle
\begingroup\renewcommand\thefootnote{\textsection}
\footnotetext{Equal contribution}
\endgroup

\begin{abstract}
The Quantum Lattice Boltzmann Method (QLBM) is one of the most promising approaches for realizing the potential of quantum computing in simulating computational fluid dynamics. Many recent works mostly focus on classical simulation, and rely on full state tomography. Several key algorithmic issues like observable readout, data encoding, and impractical circuit depth remain unsolved. As a result, these are not directly realizable on any quantum hardware. We present a series of novel algorithmic advances which allow us to implement the QLBM algorithm, for the first time, on a quantum computer. Hardware results for the time evolution of a 2D Gaussian initial density distribution subject to a uniform advection-diffusion field are presented. Furthermore, 3D simulation results are presented for particular non-uniform advection fields, devised so as to avoid the problem of diminishing probability of success due to repeated post-selection operations required for multiple timesteps.
\par We demonstrate the evolution of an initial quantum state governed by the advection-diffusion equation, accounting for the iterative nature of the explicit QLBM algorithm. A tensor network encoding scheme is used to represent the initial condition supplied to the advection-diffusion equation, significantly reducing the two-qubit gate count affording a shorter circuit depth. Further reductions are made in the collision and streaming operators. Collectively, these advances give a path to realizing more practical, 2D and 3D QLBM applications with non-trivial velocity fields on quantum hardware. 
\end{abstract}

\section{Introduction}
\begin{figure*}[t]
    \centering
    \includegraphics[width=0.94\linewidth]{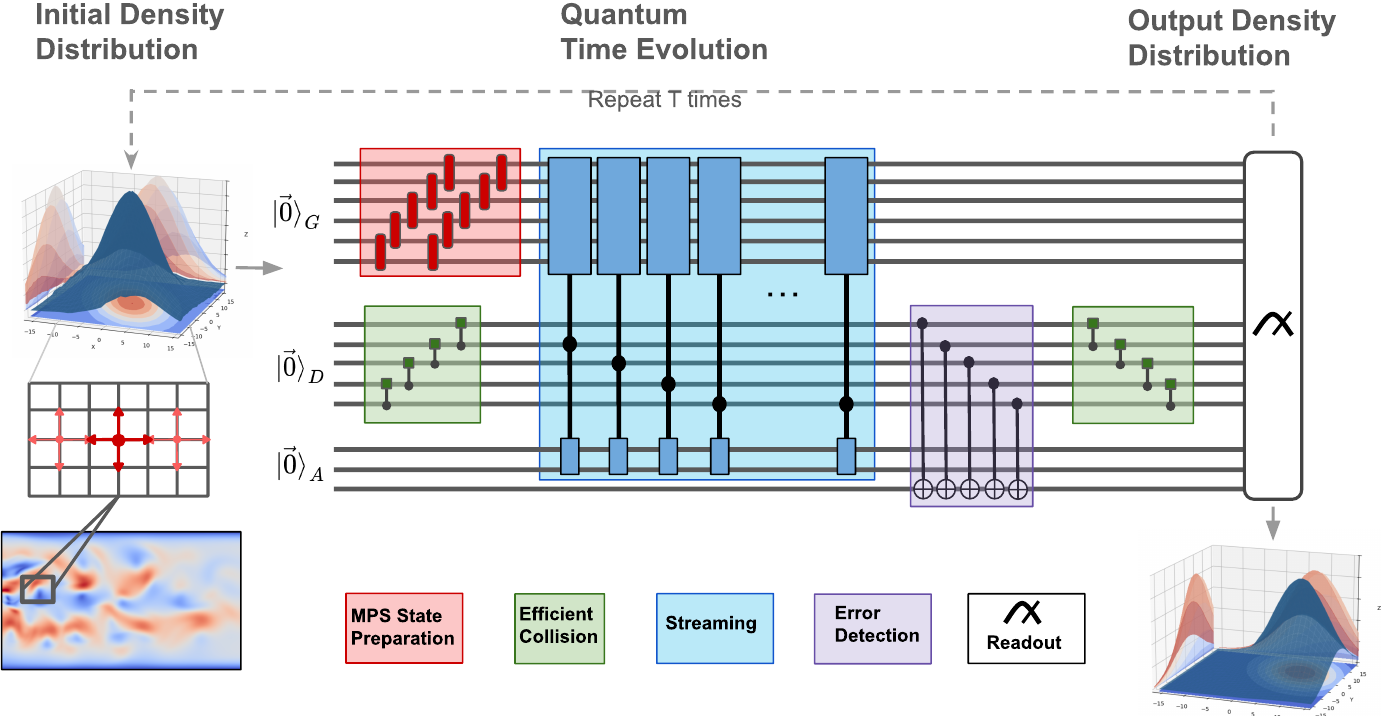}
    \caption{An overview of the full QLBM pipeline. The mesoscopic D2Q5 lattice Boltzmann formulation is used to simulate the macroscopic fluid dynamics. The quantum algorithm a smooth density distribution as input (in this case a 2D Gaussian distribution) and applies T iterations of the QLBM quantum circuit. The input distribution loaded into the amplitude of the $\log(N)$ grid qubits, $\ket{0}_G$, using a MPS state preparation circuit. The collision operator is prepared efficiently using a unary state preparation circuit on the $M$ ``direction qubits", $\ket{0}_D$, as described in the text. The streaming step acts as a series of operations on the grid qubits which are controlled on the one-hot encoded direction qubits. Additional ancilla qubits, $\ket{0}_A$, are used to reduce the total gate count of the multi-control unitary gates. Additional ancilla qubits are also used to detect certain errors which occur during the circuit execution.  }
    \label{fig: qlbm_circuit}
\end{figure*}

Numerical simulation of fluid dynamics is one of the most active areas of research. Due to the ubiquitous nature of the problem, it finds applications in diverse areas such as aeroacoustics, external/internal aerodynamics, turbulent flows etc. High-fidelity computational fluid dynamics (CFD) simulations account for a substantial share of high performance computing resources and is only expected to grow in the future \cite{slotnick2014}. The ever increasing demands push classical high performance computing to its limits, resulting in a growing interest in employing quantum computing to accelerate computation \cite{steijl2018, gaitan2020b}. The Lattice Boltzmann Method (LBM) is particularly appealing to massively parallel computation, given the localized nature of the operations involved, and the relatively simple implementation of the algorithm. The LBM views the solution at a mesoscopic scale, iteratively evolving a probability distribution function to characterize the statistical behaviour of a system of particles. The evolution involves the streaming and collision steps, which simulate the behaviour of particles over various lattice sites, followed by recovery of macroscopic quantities from the individual species, and applying boundary conditions.
A quantum algorithm to solve the collisionless Boltzmann equation was proposed in \cite
{TodorovaSteijl} and a novel streaming step was introduced in \cite{Schalkers2024} using the quantum Fourier transform. A theoretical framework for implementing the nonlinear collision operator using a Carleman linearization technique was proposed by Itani et.\ al.\  \cite{Itani2023}. Development of numerical methods in CFD typically starts with solving a model equation, for example, the advection-diffusion equation, and progressively building towards the Navier-Stokes system of nonlinear equations.
The first attempt to find quantum circuit analogues of the individual steps involved in classical LBM to solve the advection-diffusion equation was made by Budinski~\cite{Budinski2021}, and was later extended to a stream function-vorticity formulation of the Navier-Stokes equations~\cite{Budinski2022}.

While much progress has been made in the development of the quantum lattice Boltzmann approach, several obstacles still remain which have so far prevented implementations of this method on actual quantum hardware and limit its potential to be developed into  a practical CFD solver. These include practical limitations imposed by the circuit depth required to implement the streaming and collision operations, as well as theoretical limitations in implementing the state preparation and readout operations. In prior works the advection-diffusion implementation was limited to uniform velocity fields and assumed readout of the full statevector, thereby making it suitable to be run only on simulators. 
\par In this work we show how these obstacles, some identified in previous works \cite{Budinski2021} \cite{Budinski2022} \cite{wawrzyniak2025a}, can be overcome through a combination of theoretical innovation and practical engineering of the quantum circuit implementation. To demonstrate our developments, we implement the time evolution of the advection-diffusion equation under a uniform velocity field on IonQ's trapped-ion quantum computer.
\par The QLBM is an explicit iterative algorithm, and involves the following steps: encoding an initial state, collision and streaming operations, recovering macroscopic quantitites from the individual species, and implementing boundary conditions. This results in advancing one timestep, and the intermediate state thus obtained, serves as the initial state for the succeeding iteration. In this work, we improve upon recent QLBM simulation implementations \cite{Budinski2021} \cite{Budinski2022} \cite{ wawrzyniak2025a}, to make them amenable to be run on hardware. Ideally, after the initial state is loaded onto the quantum computer, it runs as many timesteps as required, and outputs an observable only after the final timestep. However, shallow circuit depths in the NISQ era necessitate running batches of fewer timesteps per circuit, reading out an intermediate state, and reloading it for the next batch of timesteps.
Given the demand to reduce circuit depth per iteration and the need to support efficient reconstruction of intermediate states by reading out appropriate observables, we introduce the main contributions of this paper:

\begin{itemize}
    \item \textit{\textbf{Encoding}}: A novel tensor network-based method to efficiently load the initial state as a matrix product state representation.
    \item \textit{\textbf{Efficient streaming operator}}: We introduce a novel implementation of the streaming operator, affording a lower circuit depth at the expense of adding more qubits, with a favourable trade-off.
    \item \textit{\textbf{Revised collision operator}}:
    We propose an LCU based approach that combines the steps of collision and computing macroscopic quantities in a way that preserves a high probability of success over repeated post-selection.
    \item \textit{\textbf{Observable readout}}: To avoid prohibitively expensive full-state tomography, we parameterize the initial condition supplied to the advection-diffusion equation, advance a timestep, and measure the parameters as output observables. This set of observables is then used to reconstruct the initial state for the subsequent iteration.
    \item \textit{\textbf{Error mitigation}}: We utilize a proposed one-hot encoding of the QLBM formulation, to detect errors. We also construct a noise estimation circuit to estimate the approximate noise channels for observable renormalization and background noise removal.
\end{itemize}

These novel algorithmic mechanisms enable an efficient realization of the QLBM for the advection-diffusion equation under a uniform field on current-generation quantum hardware. We generalize the approach to non-uniform flow fields through simulation, highlighting its potential for scalability and advancing towards relevant industrial applications.

\subsection{Background}
The evolution of a concentration $\Phi(\textbf{x},t)$ governed by linear advection-diffusion, under the action of a velocity field $\textbf{u}(\textbf{x})$, and a constant diffusion coefficient $D$, is modelled as,
\begin{equation}\label{eq:advecdiff}
    \partial_t \Phi = D\nabla^2\Phi - \nabla \left(\textbf{u}\Phi\right).
\end{equation}
Solving Eq.\ \eqref{eq:advecdiff} numerically with LBM involves solving for the macroscopic density $\Phi\left(\textbf{x}, t\right)$ as the evolution of individual species $f_i\left(\textbf{x}, t\right)$ of microscopic particle densities, such that $\Phi(x,t) = \sum_i f_i\left(\textbf{x}, t\right)$. LBM prescribes marching in time $t$ with a timestep $\Delta t$ as,
\begin{eqnarray}
\label{eq:lbm}
    f_i(\textbf{x}+c_i \Delta t, t+ \Delta t) = (1-\omega)f_i(\textbf{x},t) + \omega f_i^{eq}(\textbf{x},t),
\end{eqnarray}
where $\omega = \Delta t / \tau$ with relaxation time $\tau$, and $f_i^{eq}$ is the equilibrium density distribution given by the Bhatnagar-Gross-Krook (BGK) formulation \cite{BGK} as,
\begin{equation}
\label{eq:eq_func}
    f_i^{eq} = \omega_i \Phi\left[1+\frac{c_i\cdot u_i}{c_s^2}\right].
\end{equation}

Existing implementations of the QLBM algorithm \cite{Budinski2021, Budinski2022, wawrzyniak2025a}, set the relaxation constant $\omega=1$. This simplifies Eqs.\ \eqref{eq:lbm} and \eqref{eq:eq_func} to produce,

\begin{eqnarray}
    \Phi(x,t) &=&\sum_i k_i\Phi\left[\textbf{x}-c_i\Delta t, t-\Delta t\right], \\
    \text{where } k_i&=&\omega_i\left[1+\frac{c_i\cdot u_i}{c_s^2}\right].\label{eq:k_define}
\end{eqnarray}

Drawing upon the work in \cite{Budinski2021}, we summarize the steps in this feed-forward process, which can be simulated using the following quantum algorithm:

\begin{enumerate}
    \item \textbf{\textit{Encoding:}} $\Phi(\textbf{x},t)$ described over discrete space-time is encoded as a quantum state $|\Phi_t\rangle_G$,
    \begin{equation}
        |\Phi_t\rangle_G := \frac{1}{\|\Phi_t\|_2}\sum_\textbf{x}\Phi(\textbf{x},t)|x\rangle_G,
    \end{equation}
    where $\|\Phi_t\|_2:=\sqrt{\sum_\textbf{x}\Phi(\textbf{x},t)^2}$
    and each gridpoint $\textbf{x}$ is represented by a computational basis state $|x\rangle$. Note that the requisite notation indicating discretization is dropped for brevity. This state is stored in a quantum register ($G$) consisting of $n_G$ `grid qubits' to represent $N=L^d$ lattice sites in $d$ dimensions.
    
    \item \textbf{\textit{Collision: }} A collision operator is applied, which (in the case of a uniform velocity field) involves the preparation of a state $|k\rangle$ with the help of an auxiliary qubit, where
    \begin{equation}
        |k\rangle=\frac{1}{\sqrt{2^{n_D}}}\sum_i k_i |i\rangle_D|0\rangle_{a}+|\chi_0\rangle_D|1\rangle_a,
    \end{equation}

    for some orthogonal state $|\chi_0\rangle$.
    The state $|k\rangle$ is prepared in a quantum register ($D$) consisting of $n_D$ `direction qubits' for $M$ directions.
    
    \item \textbf{\textit{Streaming: }} Unitaries $S_i$ are constructed, such that
    \begin{equation}
        S_i|x\rangle_G=|x+c_i\Delta t\rangle_G.
    \end{equation}

    These unitaries are controlled on the direction qubits, and are applied sequentially to create a streaming operator $U_S$:

    \begin{equation}\label{eq:stream_defined}
        U_S=\prod_i \left(\mathbbm{1}_G\otimes (\mathbbm{1}-|i\rangle\langle i|)_D+S_i\otimes(|i\rangle\langle i|)_D\right).
    \end{equation}

    Applying this operator on the prepared state $|\Phi_t\rangle_G|k\rangle_D$ yields,

    \begin{align}
        \frac{1}{\|\Phi_t\|_2\sqrt{2^{n_D}}}&\sum_{x,i} k_i|i\rangle_D\Phi(x-c_i\Delta t, t)|x\rangle_G|0\rangle_a\nonumber\\
        &+|\chi_1\rangle_{D,G}|1\rangle_a,
    \end{align}

    where $|\chi_1\rangle_{D,G}$ is again some orthogonal state.

    \item \textbf{\textit{Macroscopic quantities: }} The final step involves adding amplitudes over the direction qubits. This is done by applying Hadamard gates on all the direction qubits, and yields the following state:

    \begin{align}
        &\frac{1}{2^{n_D}}\sum_x\left(\sum_i k_i\Phi(x-c_i\Delta t, t) \right)|x\rangle_G|0\rangle_{D,a}\nonumber+ |\chi_2\rangle\\
        =&\frac{1}{2^{n_D}}\sum_x \Phi(x,t+\Delta t)|x\rangle_G|0\rangle_{D,a}\nonumber+|\chi_2\rangle\\
        \label{eq:final_state_old}
        =&\frac{\|\Phi_{t+1}\|_2}{2^{n_D}\|\Phi_t\|_2}|\Phi_{t+1}\rangle_G|0\rangle_{D,a}+|\chi_2\rangle
    \end{align}

    Here, $|\chi_2\rangle$ is an orthogonal state such that

    \begin{equation}
        (\mathbbm{1}_G\otimes\langle0|_{D,a})|\chi_2\rangle=0.
    \end{equation}

    Post-selecting on the $|0\rangle_D$ state gives us the desired evolved state $|\Phi_{t+1}\rangle$.
    
\end{enumerate}

\section{Toward hardware} 

\begin{figure*}[t]
    \centering
    a) \includegraphics[width=0.63\linewidth]{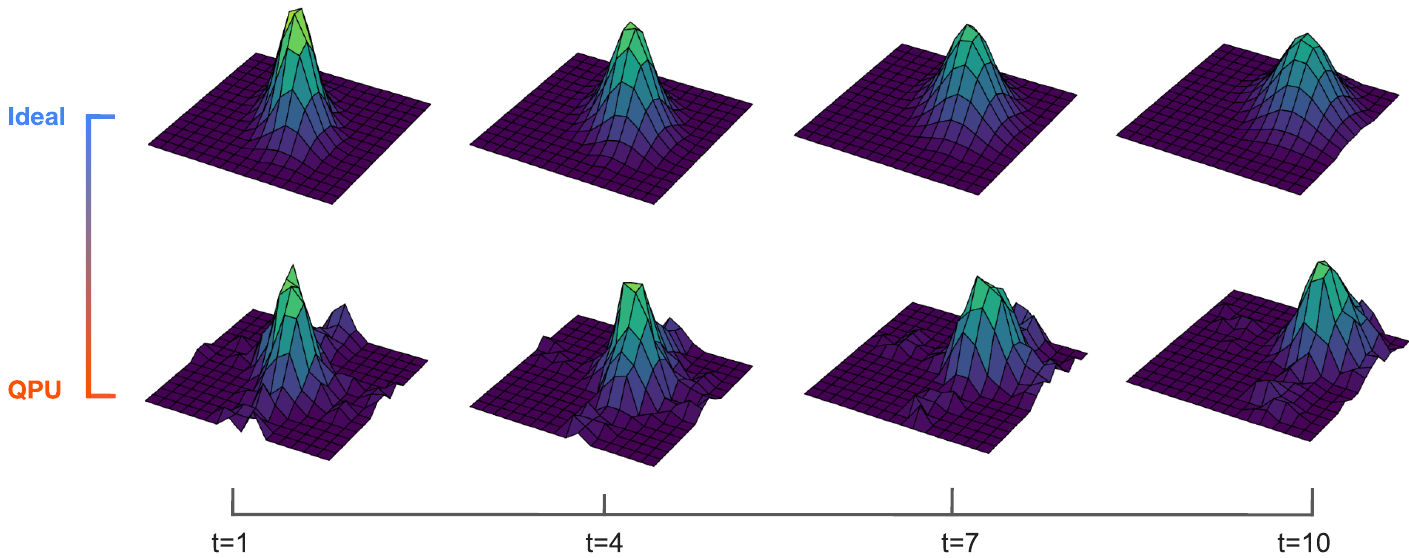}
     b) \includegraphics[width=0.30\linewidth]{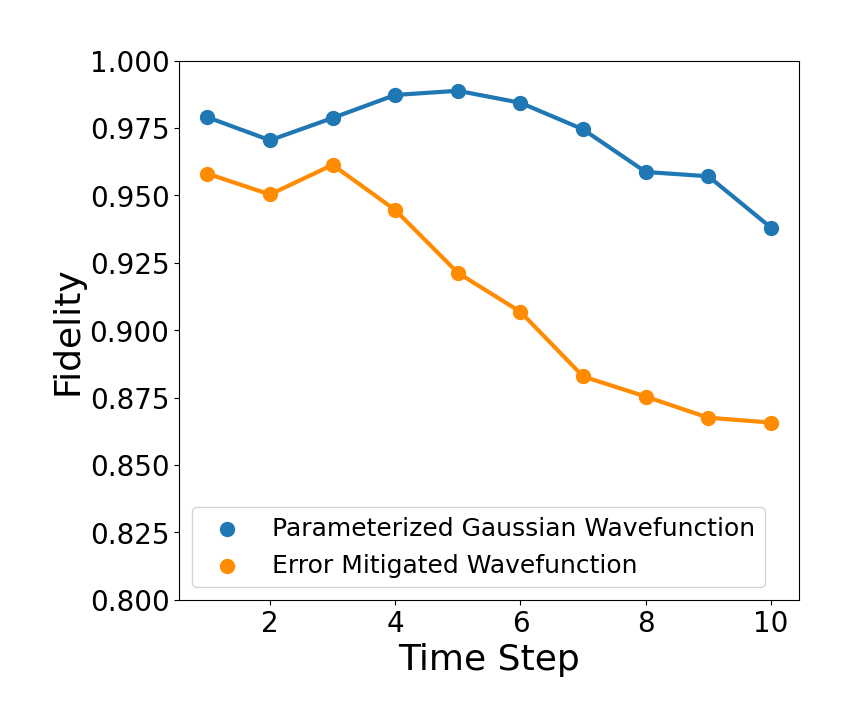}
    \caption{{\bf a)} The time evolution of a 2D Gaussian density distribution for 10 time steps using the QLBM algorithm evaluated using an ideal simulation (top), and the error mitigated wave function executed on the IonQ Forte quantum computer (bottom). {\bf b)} The fidelity, $|\langle \psi | \phi\rangle|$, between the density distribution measured from the QPU and the classical LBM solution (orange curve). This fidelity is further improved by reconstructing a 2D Gaussian distribution using observables measured from the wave function, where over $92\%$ fidelity is achieved at all time steps (blue curve).  }
    \label{fig:qpu_results}
\end{figure*}

There are at least three major obstacles with current approaches to the quantum lattice Boltzmann method which must be overcome in order to implement the algorithm on NISQ devices. These are: preparation of the initial density function, the high circuit depth needed to implement the streaming and collision operators, and the inefficient readout of the intermediate and final states throughout the evolution. The state preparation and readout operations must be performed in a time which scales polynomially in the number of qubits (and logarithmically in the number of lattice grid points), in order for there to be any potential for a quantum speedup using this method. It is well known that preparing an arbitrary initial quantum state using a quantum circuit generally requires a circuit whose depth scales exponentially in the number of qubits. While the initial states required for the QLBM may be far from an arbitrary quantum state, typical methods are only able to prepare very simple initial distributions, strongly limiting the scope of simulations which can be performed \cite{wang2025}. 

\subsection{Hardware results}
In this work, we implemented the QLBM algorithm on hardware, to solve the advection-diffusion equation for a 2D initial Gaussian density distribution on a $16\times 16$ grid using the IonQ Forte trapped-ion quantum computer. The quantum circuit used $19$ qubits and $\sim 260$ two-qubit gates. The results of this execution are shown in Fig.~\ref{fig:qpu_results}, and demonstrate agreement with the classical LBM simulation. 

In order to execute the algorithm, the initial density distribution at each time step is loaded into the quantum circuit using an MPS circuit approximation to the exact distribution. The time evolution is then carried out by applying novel improved versions of both the collision and streaming operators from ref.~\cite{Budinski2021}. We bypass the need to perform full quantum state tomography at each time step by measuring a few observables which parameterize the time evolved Gaussian distribution. This distribution is then reloaded into the quantum circuit for each step of the time evolution.  In Fig.~\ref{fig:qpu_results} a), we show the density measured from the time-evolved state after error mitigation is applied to the wave function. The fidelity of the exact solution with both the error mitigated wave function and the parameterized Gaussian state with the parameters measured from this wave function are shown in Fig.~\ref{fig:qpu_results} b), where we achieve over $92\%$ fidelity at all time steps, and over $97\%$ fidelity for the first seven.

Further details of novel advancements in the implementation which allowed us to obtain these results are described in the remainder of this section, while Sec.~\ref{sec:error_mit} is dedicated to describing the error mitigation methods used to achieve optimal performance on the QPU hardware.

\subsection{MPS Loading}\label{sec:mps}
When the amplitudes of a quantum state can be represented by a smooth function over a binary encoding of the basis states, the entanglement of the state increases very gradually as the number of qubits grow. The authors in Ref.~\cite{matsuura_ieee_2020, ran_pra_2020}, leveraged this property to develop an efficient technique for approximating probability distributions using MPS and quantum circuits with low depth. A MPS state of bond dimension $\chi = 2$ can be represented exactly with a single layer of two-qubit gates. For more complicated functions,
the entanglement structure becomes richer, increasing the MPS bond dimension $> 2$ and also the number of layers of two-qubit gates. In Refs.~\cite{iaconis_johri_mps_2024,iaconis2023tensor}, MPS methods are used to efficiently represent smooth differentiable probability distributions like normal distributions, which can be encoded iteratively onto quantum circuits by adding layers of two-qubit entangling unitary gates.

We use this MPS technique for state preparation in the QLBM algorithm, using 2-3 layers of two qubit gates, depending on a tradeoff between state encoding accuracy and optimizing the number of 2-qubit gates to limit hardware noise. Using a Gaussian initial state, the amount of entanglement remains low during the advection-diffusion evolution and the approximation accuracy of the MPS method remains high. The MPS loading technique requires $2 \mathrm{n_{layers}} (n_G-1)$ CNOTs compared to conventional encoding techniques like amplitude encoding which require $\sim O(2^{n_G})$ CNOTs, where $\mathrm{n_{layers}}$ is the number of 2-qubit gate layers. For the final implementation on IonQ Forte, when we run the full algorithm with 10 steps of evolution, we use 2 such MPS layers, requiring $28$ CNOTs for state preparation. 

We perform observable readout and the state preparation (MPS loading) circuit after every time step to limit the the hardware implementation to low-depth circuits. The full circuit, including MPS state preparation, streaming and direction preparation modules of the QLBM method, and readout, required $\sim 250$ CNOTs at each step of the evolution.

\subsection{Increase the probability of success at postselection}\label{sec:improve_prob_success}
From Eq. \eqref{eq:final_state_old}, we see that the probability of successful postselection is given by

\begin{equation}
    \text{Pr}(|0\rangle_D)=\frac{\|\Phi_{t+1}\|^2}{2^{2n_D}\|\Phi_t\|^2}.
\end{equation}

In this section, we describe how we modify the algorithm to increase this probability to

\begin{equation}\label{eq:new_success_prob}
    \text{Pr}(|0\rangle_D)=\frac{\|\Phi_{t+1}\|^2}{\|\Phi_t\|^2}.
\end{equation}

We motivate this by recognizing that this algorithm aims to take a quantum state $|\Phi_t\rangle$ and compute the state

\begin{equation}\label{eq:LBM_LCU_op}
    \frac{\|\Phi_{t+1}\|}{\|\Phi_t\|}|\Phi_{t+1}\rangle=\sum_i k_iS_i|\Phi_t\rangle.
\end{equation}

Here, $S_i$ is a unitary operation and $k_i$ are positive real numbers such that $\sum_i |k_i|=1$.


We want to create a circuit that applies the operation in Eq. \eqref{eq:LBM_LCU_op}. A natural framework to do so is Linear Combination of Unitaries (LCU) \cite{lcu1,lcu2}. Given unitary operators $\{U_i\}_i$ and coefficients $\{\alpha_i\}_i$, the LCU method can be used to construct an operator $U_{LCU}$ such that

\begin{equation}
\left(\mathbbm{1}\otimes\langle0|_a\right)U_{LCU}\left(|\psi\rangle\otimes|0\rangle_a\right)=\frac{1}{\sum_j|\alpha_j|}\sum_i\alpha_iU_i|\psi\rangle.
\end{equation}

We can use this method to create the full circuit $U$. Here we have our unitary operators $\{S_i\}_i$, and their associated coefficients $\{k_i\}_i$ -

\begin{equation}
    \langle0|_DU|0\rangle_D=\sum_i k_iS_i
\end{equation}

Applying this circuit on the initial state gives us

\begin{equation}\label{eq:final_LCU_state}
    U|\Phi_t\rangle_G|0\rangle_D=\frac{\|\Phi_{t+1}\|}{\|\Phi_t\|}|\Phi_{t+1}\rangle_G|0\rangle_D+|\chi\rangle,
\end{equation}

where $|\chi\rangle$ is a garbage state such that $(\mathbbm{1}_G\otimes\langle0|_D)|\chi\rangle=0$.


From here, we see that the probability of finally measuring the direction qubits to be in the $|0\rangle_D$ state has indeed increased to the larger value given in Eq. \eqref{eq:new_success_prob}. This improvement has further implications that will be discussed in Section \ref{sec:time_comp}.

While this new method using the LCU approach is similar to the existing implementation, there are a few key differences:

\begin{enumerate}
    \item For the collision step, the previous method prepares the state $|k\rangle_D$ defined in Eq. \eqref{eq:k_define} using a method that involves post-selection of an auxiliary qubit with a success probability of $1/2^{n_D}$. While the new method also involves state-preparation on the direction qubits, we use a state preparation routine  that doesn't require a post-selection step.
    \item In the collision step, instead of preparing the state $|k\rangle$, the LCU approach uses a standard state preparation routine to create the `PREP' operator \cite{johri2021nearest}  -
    \begin{equation}
        \text{PREP}|0\rangle_D=\sum_i \sqrt{k_i}|i\rangle_D.
    \end{equation}
    \item For the final step in the existing implementation, Hadamard gates are applied to the direction qubits. In the LCU approach, we replace the final step with an application of PREP$^\dagger$ on the direction qubits.
\end{enumerate}

\subsection{Improved Streaming Operator}\label{sec:improve_streaming}

The typical implementation of the QLBM algorithm encodes the density function $f_i(x)$ using $\log_2(N)$ grid qubits and $\lceil\log_2(M)\rceil$ direction qubits. In this work, we develop a novel implementation of the direction qubits which greatly reduces the number of overall two-qubit operations which must be performed, at the expense of moderately increasing the required number of qubits needed. This favorable depth vs width trade-off, along with the other improvements described in this work, bring the required resources for executing the QLBM algorithm down to a scale in which we can run a highly nontrivial demonstration on current quantum computing hardware. 
The main insight is the realization that a large number of controls can be removed from the streaming operator by using an alternate encoding on the direction qubits. 

\begin{figure}
    \centering
    \hspace{-5mm} \includegraphics[width=0.75\linewidth]{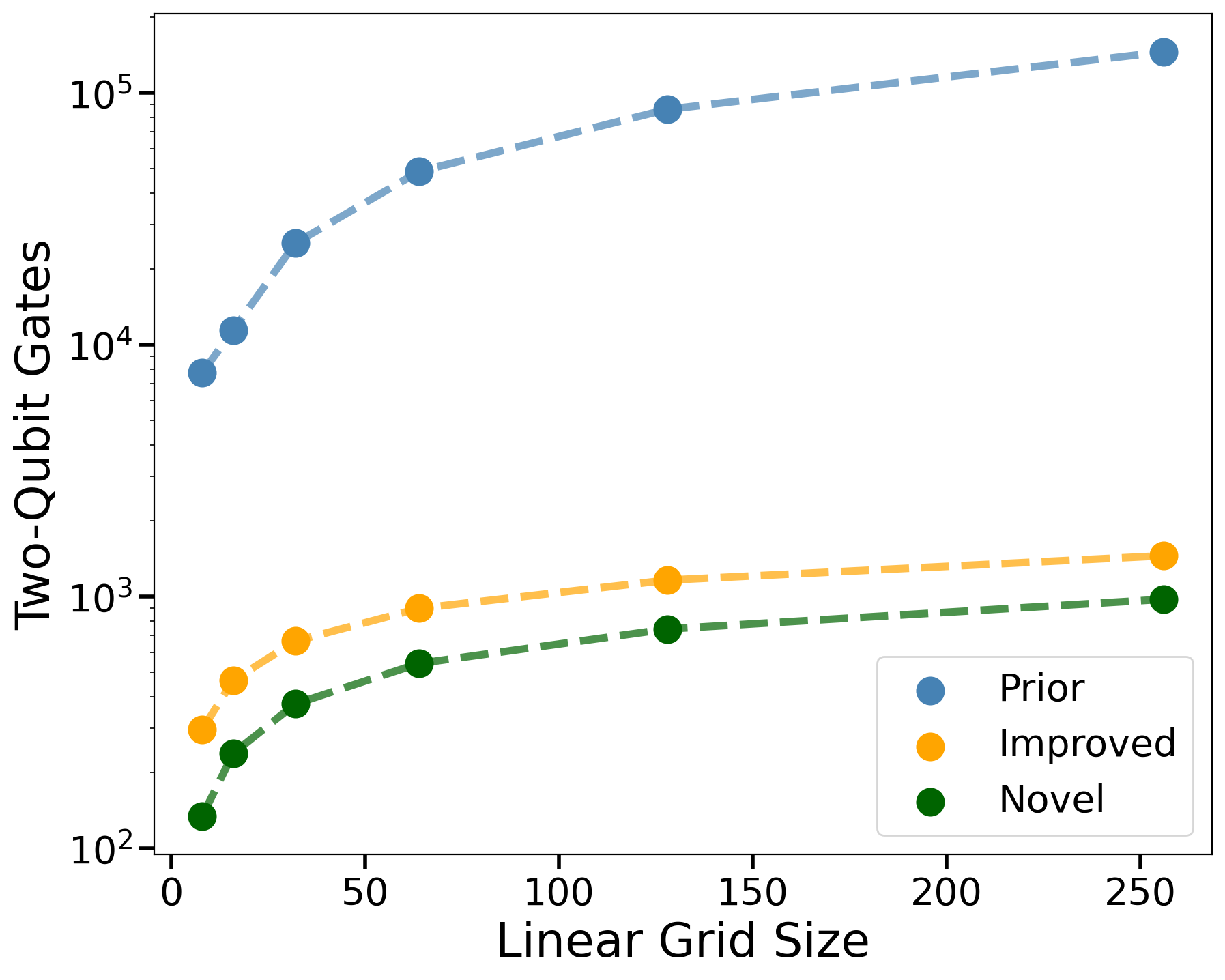}
    \caption{The two-qubit gate count of the streaming operator as a function of lattice grid size for the D2Q5 advection-diffusion model using the prior compilation technique from literature, the improved compilation with ancilla qubits, and using our novel one-hot encoding scheme.}
    \label{fig:stream_gate_count}
\end{figure}

Within the QLBM algorithm, amplitudes of the the direction qubits, $\ket{\psi}_D$, encode the propagation velocities along the $M$ directions of the LBM model. The streaming of the density function, $f_i$, in each direction is then performed by applying a separate unitary operation which is controlled on each of the $M$ direction states encoded in $\ket{\psi}_M$. The streaming operator itself contains multi-control Toffoli gates, C$^\text{n}$X, which are used to implement a cyclic permutation operator along a particular lattice direction. Quantum computers typically do not natively implement arbitrary unitary operations, but instead decompose complex unitaries into a series of basic one and two qubit operations. A naive decomposition of the C$^\text{n}$X operator uses $\mathcal{O}(n^2)$ two qubit gates \cite{nielsen2010quantum}, leading to intractable circuit depths when implementing the QLBM algorithm, as shown in Fig.~\ref{fig:stream_gate_count}. However, large reductions in this circuit depth can be achieved using ancilla qubits in this decomposition, where a C$^\text{n}$X gate can be implemented using as few as $\mathcal{O}(6n-12$) CNOT gates \cite{dmitri2015advantages}.

In our work, instead of using $\lceil \log(M) \rceil$ qubits to encode the M directions of propagation, the major innovation is to use a one-hot encoding of the direction qubits which use $M$ qubits, so that
\begin{eqnarray}
    \ket{\psi}_D &=& \sum_i \sqrt{k}_i \ket{e_i} \\
   \ket{ e_i} &=& \ket{0\dots 010 \dots 0}
\end{eqnarray}
where $\ket{e_i}$ is a Hamming weight 1 basis state with a single 1 on site i. If we then apply the controlled streaming operations using this one-hot encoding formulation, each streaming unitary is always controlled on just a single direction qubit, regardless of the number of streaming directions in the model. Applying the streaming unitary $U_i$, controlled on the i$^\text{th}$ direction qubit then results in the entangled state
\begin{eqnarray}
    \ket{\Psi_{t+1}}= \sum_i \sqrt{k}_i \ket{e_i}U_i \ket{\psi_t}
\end{eqnarray}
as desired. 

This one-hot encoding of the direction qubits removes $(\lceil \log_2(M)\rceil -1)$  controls on each unitary gate in the streaming operator, resulting in a savings of $\frac{6}{2}M(\lceil\log_2(M)\rceil-1)\log_2(L)(\log_2(L)-1)$ CX gates for the entire streaming operator, compared to improved implementation with densely encoded direction qubits (which also uses ancilla qubits to implement to C$^{\text{n}}$X operators). As shown in Fig.~\ref{fig:stream_gate_count}, for the D2Q5 model on a $16\times16$ lattice, which we implement on the IonQ quantum computer in this work, this results in a nearly $ 2\times$ improvement in two-qubit gate count. 

In Table~\ref{tab:cx_count}, we show that the relative improvement in gate count offered by the one-hot encoding scheme increases for more complex QLBM models in higher dimensions and with additional lattice directions, due to the increased number of direction qubits needed to construct the corresponding circuits. Further improvements may be expected for even more complex cases such as the streaming-vorticity model or models using Carleman linearization, where a large number of additional degrees of freedom are needed for the implementation.

\begin{table}
    
\begin{center}
\begin{tabular}{c c c c c c} 
 \hline\hline\\[-1ex]
 Model & \shortstack{Grid\\Size} &\shortstack{Stand.\\ Qubits}
 &\shortstack{Stand. \\ CX Count}& \shortstack{Novel \\ Qubits} & \shortstack{Novel \\ CX Count} \\ [0.5ex] 
 \hline
 D2Q5* & $16^2$ & 15 & 480 & 15 &244  \\[-0.3ex]
 D2Q5 & $32^2$ & 18 & 672 & 18 &388 \\
 D2Q5 & $1024^2$& 33 & 1992 & 33 &1468 \\
 D2Q9 & $16^2$ &17 & 1152 & 19& 488 \\
 D2Q9 & $32^2$ &19 & 1824 & 21 & 776 \\
 D3Q19 & $32^3$ & 25 & 4104 & 37& 1746 \\
 D3Q27 & $32^3$ &25 & 5928 & 45&2522 \\ 
 D3Q27 & $1024^3$ &47 & 16068 & 65 & 7826  \\ [1ex] 
 \hline\hline
\end{tabular}
\end{center}
\caption{Two qubit gate counts of the streaming operator using the standard (improved) decomposition with ancilla qubits, and the novel implementation used in this work for different grid sizes and models. $*$ indicates the model executed on the IonQ QPU backend in this work.}\label{tab:cx_count}
\end{table}

\subsection{Observable readout}
The exponential growth of resources in full quantum state tomography and classical post-processing, especially when scaled to large qubit systems, pose a key bottleneck in realistic and useful implementation of a majority of quantum algorithms \cite{nielsen2010quantum}. Even if advantage is gained through a quantum algorithm during the computation phase, when we readout a large vector element-wise (especially when scaling to large sizes), the quantum advantage will be lost if we need to read out and store all vector elements. 

In this work, to preserve the advantage and bypass this bottleneck in the QLBM pipeline, we parametrize the Gaussian wave function using the mean, variance and covariance, and measure only these observables at every evolution step. We reconstruct the corresponding wave function using these measurements and load the state for the next time step. This approach enables us to run a fixed number of circuits ($\sim 1000 -10000$ shots) and store only 3 values at each step to approximate and reconstruct the full state, bringing us one step closer to solving such time evolution problems on a real quantum device with favorable resource scaling. Since the initial state is Gaussian and remains approximately so during the evolution we study here, these 3 observables exactly (or very close to exact) parametrize the full state. This technique of using observables for state reconstruction can be applied to other types of smooth initial states as well. For example, for a state that is not purely gaussian, one can also include skewness or kurtosis along with the above set of observables to approximate state reconstruction to some degree of accuracy. For more complicated states, other possible reconstruction techniques like neural network assisted learning of local regions in a wave function ~\cite{PRR_2021} or using local observables for density matrix reconstruction either through maximum likelihood estimation~\cite{ mle_PRA_1999, mle_Baumgratz_2013, robbin_prl_2010, Singh_2016} or through iterative training methods like matrix product state approximations or neural networks~\cite{20q_recon_PRXQ_2023, 20q_state_entanglment_blatt_2018, torlai_melko_PRL_2019} can also be used and integrated with the QLBM pipeline.

\begin{figure}
    \centering
    \includegraphics[width=1.0\linewidth]{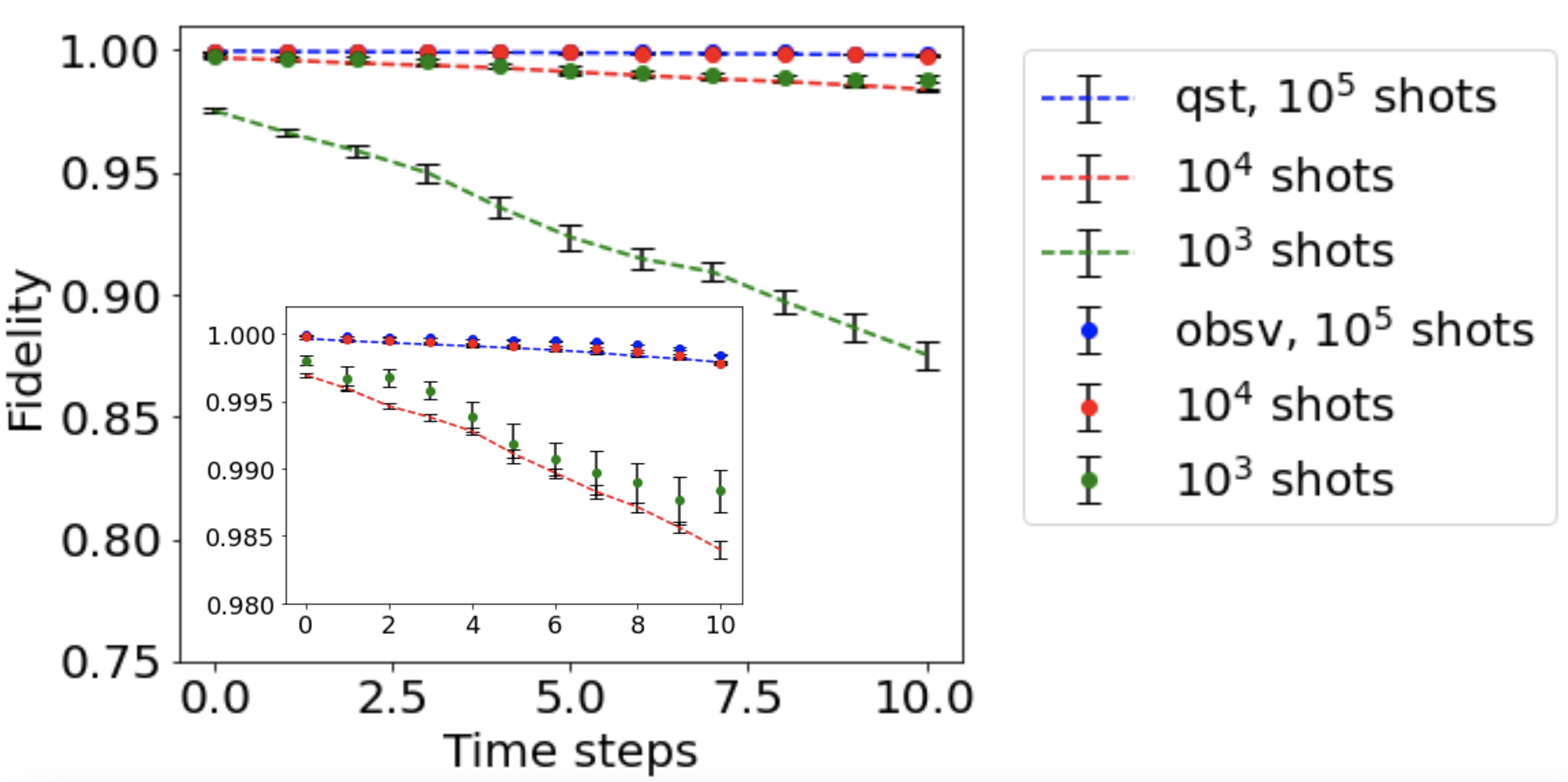} 
    \caption{We show the fidelity of the quantum states obtained from the QLBM circuit using shot based simulator and compare between full state tomography (qst) and observable based reconstruction (obsv). Fidelity is calculated based on the ideal simulation of the above strategies and the output of statevector simulation.}
    \label{fig: qst_vs_obst}
\end{figure}

\section{Error Mitigation}\label{sec:error_mit}
\subsection{Error Detection using Flag Qubits}

The one-hot encoding formulation of the QLBM algorithm also offers several practical advantages in terms of error mitigation and error resilience of the algorithm. We are able to detect a portion of the errors during the circuit execution using the error detection module shown in Fig.~\ref{fig: qlbm_circuit}. Since the direction qubits must always remain in the Hamming weight $=1$ subspace, any operation which takes the state out of this subspace must have been caused by an error. The quantum circuit used to flag ``weight one" errors in $\ket{\Psi}_D$ simply consists of a series of CX gates for each direction qubit with the target on a shared ancilla qubit. This effectively measures the parity of the state on the direction qubits. Any error which changes the Hamming weight of the direction qubits by one will flip the state of the ancilla qubit and can be detected and discarded when all qubits are measured at the end of the circuit. This error detection step must be performed before the final $\text{PREP}^\dagger$ operator is applied on the direction qubits, which rotates the one-hot encoded subspace on the direction qubits out of the Hamming weight $=1$ subspace into a superposition over the ``good" all-zero state, $\ket{0}_D$ and orthogonal state $\ket{\perp}_D$. This post-selection process then removes shots in which errors occurred, at the cost of reducing the total number of usable shots in the good subspace that can be obtained in a given time. In our experiments on the IonQ Forte QPU, we applied this error mitigation circuit 4 times, once after after each applications of the streaming operator for a particular direction, with the target on a separate ancilla qubit each time. Note that Fig.~\ref{fig: qlbm_circuit} includes only the last application of this module.  We additionally apply post-selection on the ancilla qubits which are used to implement the multi-control unitary gates, as we know that these qubits should remain in the zero state at the end of the circuit execution. Finally, we note that the one-hot encoding also has a built in noise resilience mechanism, as some errors which occur on the direction qubits will automatically be transformed into the perpendicular subspace after the final $\text{PREP}^\dagger$ unitary operator is applied. Since the $\text{PREP}$ circuit will only transform Hamming weight $=1$ states to the all zero state, any errors in $\ket{\Psi}_D$ at the end of the streaming step will be mapped to the perpendicular subspace. Together, these facts allow us to extract a useful signal from the noisy hardware despite the high complexity of the quantum circuit.

\begin{figure}[t]
    \centering
    \includegraphics[width=\linewidth]{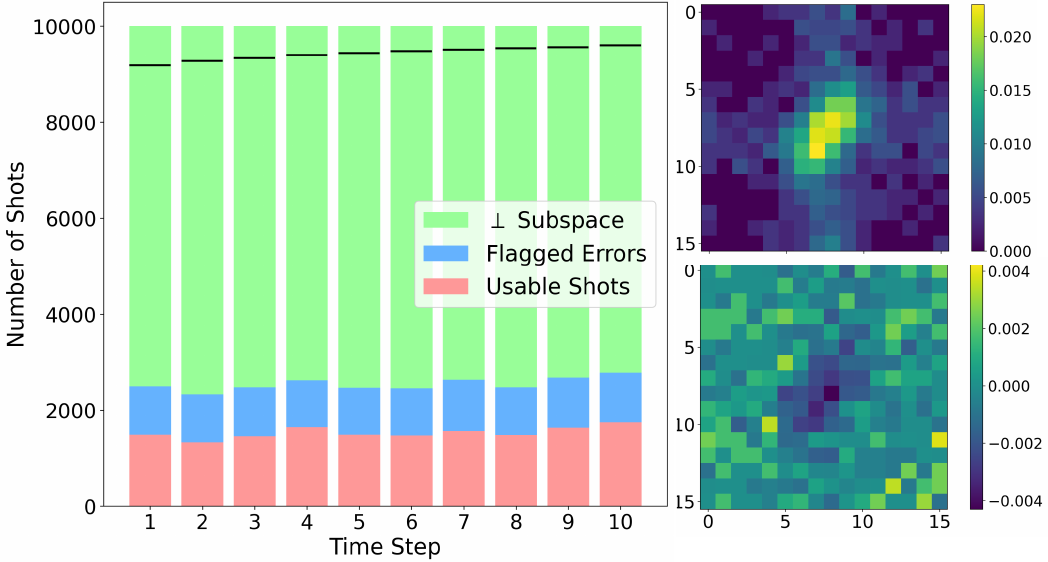}
    \caption{{(\it left)} The share of total shots measured from the QPU backend for each time-step, which are in: the perpendicular subspace (green), the good subspace but with flagged errors (blue) and the good subspace with no detected errors (red). Black bars show the number of usable shots in the ideal noiseless case. {\it(upper right)} The density distribution after one timestep using only the usable shots. {\it (lower right)} The difference in the density before and after the flagged errors are removed.   }
    \label{fig:error_shots}
\end{figure}

In Fig.~\ref{fig:error_shots}, we show the number of errors which occurred for each of the 10 time steps in our QPU simulation, where 10000 shots are taken at each time step. The black bars indicate the number of usable shots in the good subspace in the ideal case where no errors occur. The green portion of each bar indicates the number of shot which are mapped to the perpendicular subspace. Out of the remaining shots which are mapped to the good subspace, the blue portion are additional measurements where errors are detected using the circuit described above. The remaining red portion represents the final measurements where no detectable errors occurred, and which can be used to calculate the output density. This ranges between 1332 and 1782 recovered shots for the 10 circuits which we evaluated on the QPU. We also show a heatmap of the density distribution after one timestep after the detected errors are removed but before any further error mitigation is applied. In the bottom right, we show the difference in the density distribution before and after the flagged errors are removed. This demonstrates that the measurements where errors are detected are essentially uniformly distributed throughout the entire lattice and therefore the errors which occur are largely correlated between the lattice qubits and the grid qubits, since removing errors detected in $\ket{\Psi}_D$ improves the distribution in $\ket{\Psi}_G$.

\subsection{Empirical noise estimation}
In Ref.~\cite{noise_estimation_circ_prl_2021}, the authors develop a method where a noise estimation circuit with similar structure as the target circuit is constructed to measure the depolarizing noise rate and correct the output of the target circuit using the measured rate. They experimentally demonstrated that the full pipeline, including the readout correction, compilation, mitigation with these noise estimation circuits and zero noise extrapolation, was able to generate very accurate results.

\begin{figure}[htbp]
    \centering
    \includegraphics[width=0.67\linewidth]{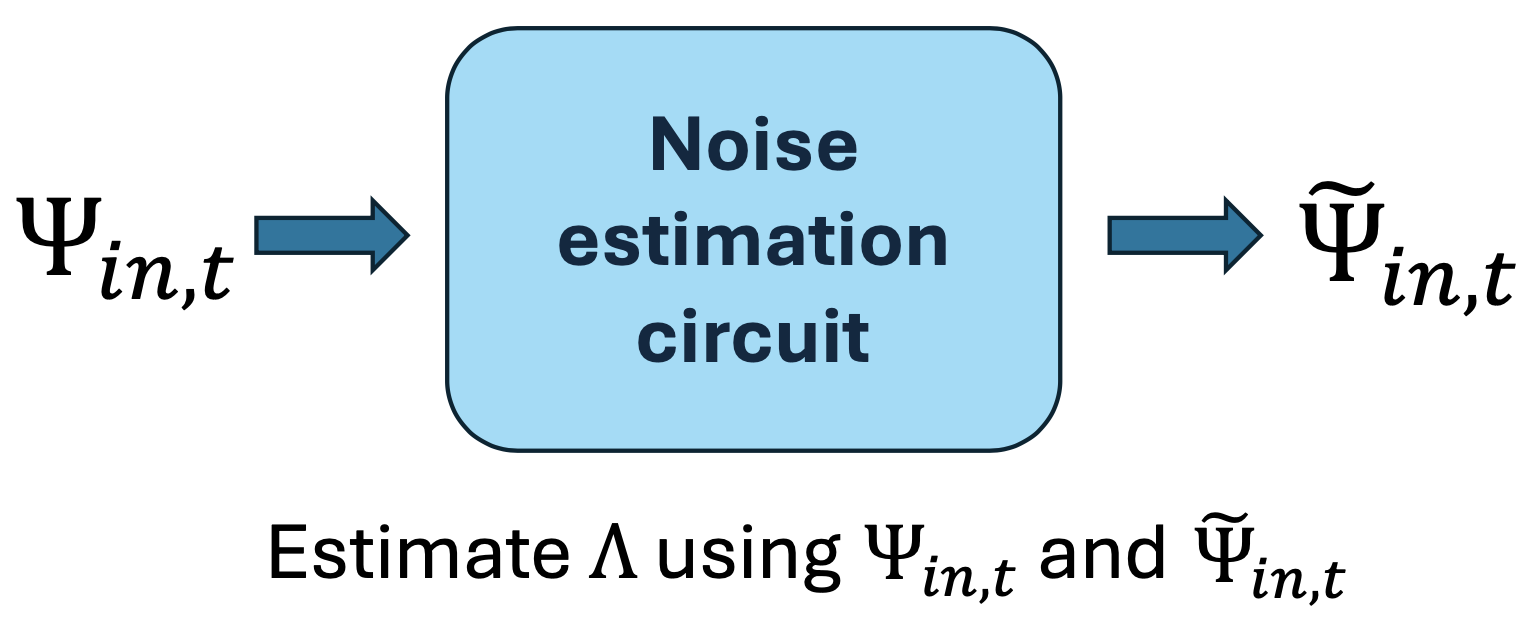} \\
    \vspace{0.5cm}

    \hspace{0.3cm}
    \includegraphics[width=0.8\linewidth]{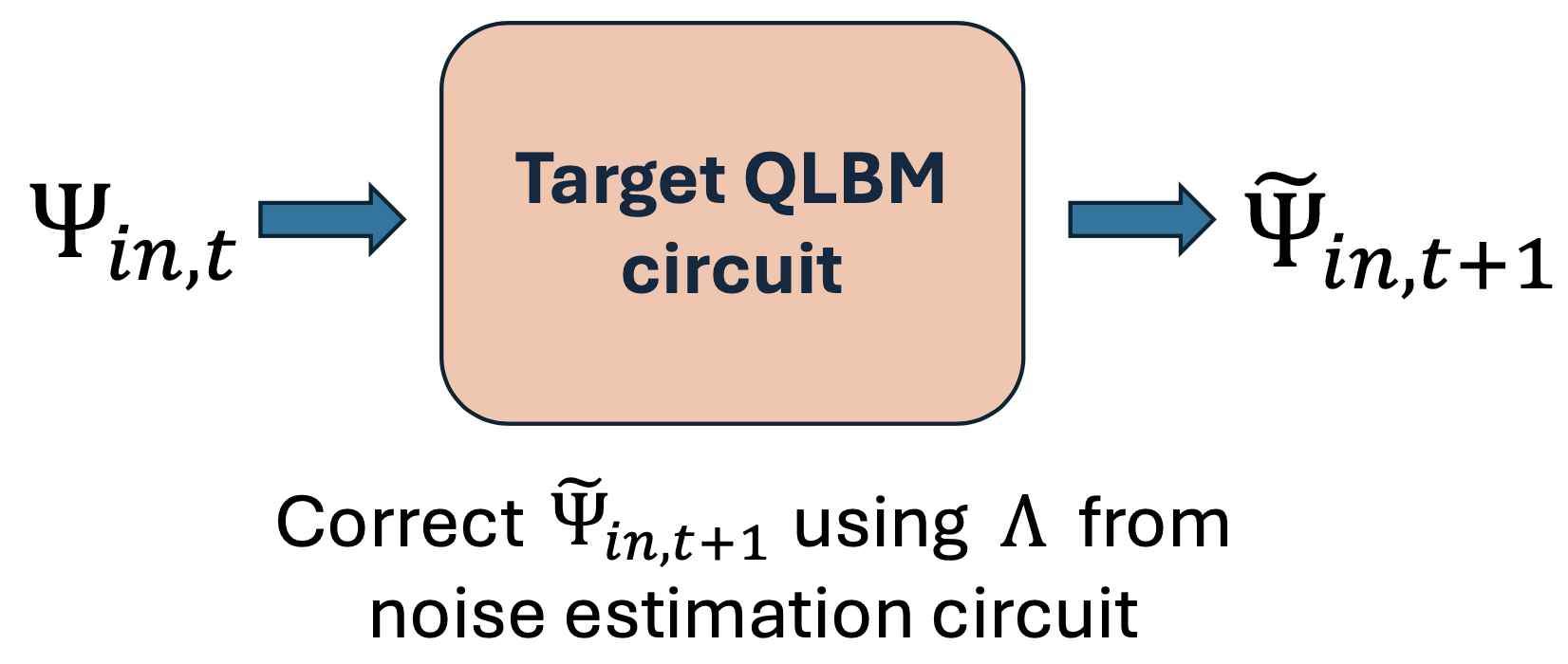}
    \caption{The error estimation procedure for one time step. We estimate the degree of noise using a circuit implementing the identity operator, and use this estimate to correct the output of the target QLBM circuit. }
    \label{fig:noise-estimation-circ}
\end{figure}

Following a similar strategy (see Fig.\ref{fig:noise-estimation-circ}), we construct a noise estimation circuit with similar structure as the QLBM algorithm, by removing the collision unitaries (the green boxes in Fig.~\ref{fig: qlbm_circuit}). Since the CNOTs in the streaming operators now act on all `0's in the absence of the collision unitaries, the estimation circuit essentially becomes close to an identity implementation circuit on the state preparation module. This ensures that the output from this estimation circuit should be the exact output of the state preparation module in the ideal case. 

Comparing the readout of the observables mean, variance and covariance from the estimation circuit on the QPU and the ideal simulation case, we can estimate the noise using a depolarizing channel model. Once the noise parameters are estimated, we can map back the noisy readout values from the target circuit to low noise or noise-free outputs and thus retrieve ideal (or close to ideal) results.

The depolarizing noise model is given by:

\begin{equation} \label{eq: depol_channel}
    \epsilon(\rho) = (1-\lambda) \rho\ +\ \lambda \frac{I}{2^n},
\end{equation}
where, $\epsilon$ is the noise channel, $\rho$ is the density matrix, $\lambda$ is the noise parameter which depends on the circuit and the device, $n$ is the number of qubits. Expectation values of Hermitian operators $\hat{O}$ also follow the same equation under the same noise channel,

\begin{equation} \label{eq: obs_noisy_channel}
    \langle \hat{O}(\epsilon)\rangle  = (1-\lambda) \langle \hat{O}_\text{ideal} \rangle\ +\ \frac{\lambda }{2^n} \langle \hat{O}\ \hat{I}\rangle,
\end{equation}

Using Eqns.\ ~\eqref{eq: depol_channel}, ~\eqref{eq: obs_noisy_channel} and the readouts from the noise estimation circuit in the noisy and ideal case, we can estimate the parameter $\lambda$. From mean and covariance, $\lambda = \frac{\langle\hat{O}_\text{ideal}\rangle - \langle\hat{O}(\epsilon)\rangle}{\langle\hat{O}_\text{ideal}\rangle }$. From variance, $\lambda = \frac{\langle\hat{O}_\text{ideal}\rangle - \langle\hat{O}(\epsilon)\rangle}{\langle\hat{O}_\text{ideal}\rangle - \langle \hat{O} \hat{I}\rangle}$.

Note that since the noise in a QPU is not an exact depolarizing model, the $\lambda$s estimated from the mean, variance and covariance are not exactly equal. Using these different $\lambda$s, we renormalize the corresponding noisy readouts with an inverse depolarizing map to estimate the error-free ones. Once these clean observables are retrieved, we can reconstruct the full quantum state by using the mitigated mean, variance and covariance. 

\subsection{Background noise removal}
Since the depolarizing noise channel is not a complete representation of all the noise sources in a real quantum device, only using the observable renormalization technique from the noise estimation circuits does not remove all the major noisy contributions {\cite{maunz_sandia_drift_2020, Majumder2023characterizing}}. From the output of the noise estimation circuit, we observe a constant background noise which pushes smaller values in the function to higher values, meaning the shots with very low probabilities in the ideal case start appearing with higher probabilities. To correct this error, we estimate a constant offset empirically from the noise estimation circuit and apply it to the output of the target circuit. Combining this background noise removal with the observable renormalization and debiasing, we are able to implement the full QLBM algorithm on IonQ Forte with considerable accuracy. 

\section{Executing multiple timesteps in a single circuit }\label{sec:time_comp}

\begin{figure}
\centering
  \includegraphics[width=0.8\linewidth]{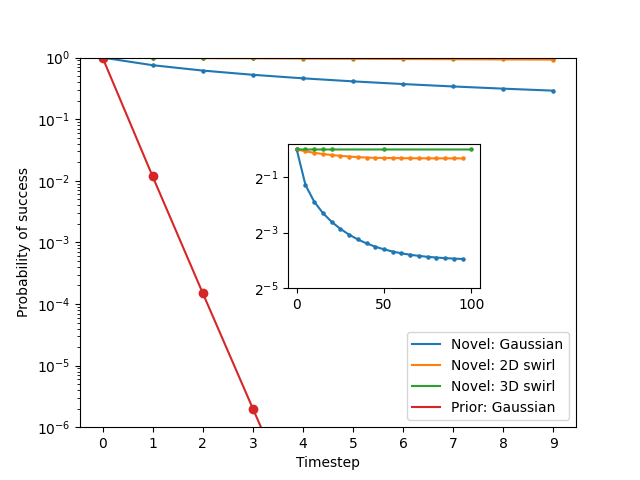}
\caption{Overall probability of success of the algorithm (with multiple timesteps in a single circuit) plotted against total number of timesteps for various cases - 2D gaussian initial state undergoing uniform advection and diffusion, using both our LCU approach (blue) and the prior implementation (red); 2D sinusoidal initial state as seen in Eq.\ \eqref{eq:sin2d} under the velocity field in Eq.\ \eqref{eq:swirl2d} (orange); $3D$ sinusoidal initial state as prescribed in Eq. \eqref{eq:sin3d} under the velocity field in Eq.\ \eqref{eq:swirl3d} (green).} \label{fig:succ_prob}
\end{figure}

So far, we have worked with a density function that maintains a parametric form throughout the evolution. We take advantage of this fact for reading out the statevector at every timestep. We can, however, bypass the need for tomography at every step by applying all the timesteps in one circuit. We denote the combined operation of measurement and post-selection of the direction qubits with a projection operator $\Pi_0:=|0\rangle_D\langle0|_D$.
Application of this operator to the final state from Eq.\  \eqref{eq:final_LCU_state}, gives,

\begin{equation}
    \Pi_0U|\Phi_t\rangle_G|0\rangle_D=\frac{\|\Phi_{t+1}\|}{\|\Phi_t\|}|\Phi_{t+1}\rangle_G|0\rangle_D.
\end{equation}

Instead of stopping to read the state $|\Phi_{t+1}\rangle_G$, we re-apply these operators to get $|\Phi_{t+2}\rangle_G$, and so on. Thus,

\begin{equation}
    \left(\Pi_0U\right)^T|\Phi_0\rangle_G|0\rangle_D=\frac{\|\Phi_{T}\|}{\|\Phi_0\|}|\Phi_{T}\rangle_G|0\rangle_D .
\end{equation}
Here, application of the projection operators require applying mid-circuit measurements on the direction qubits. If any of the T measurement rounds return a state other than $|0\rangle_D$, the full circuit needs to be repeated until success. From Eq.\ \eqref{eq:final_LCU_state}, we see that the probability of success for the $i^{\text{th}}$ measurement is given by $\text{Pr}(|0\rangle_D)={\|\Phi_{i+1}\|^2}/{\|\Phi_i\|^2}$.

Multiplying these probabilities over all timesteps gives the overall probability of success for the full algorithm -

\begin{equation}
    \text{Pr}(|0\rangle_D)=\frac{\|\Phi_{T}\|^2}{\|\Phi_0\|^2}.
\end{equation}

\begin{figure*}
\centering
\subfloat[$T=0$]{
  \centering
  \includegraphics[width=0.25\linewidth]{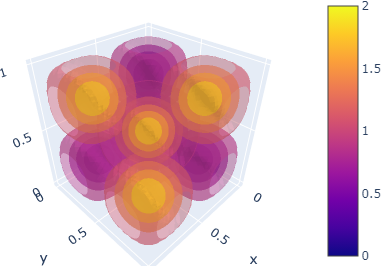}
}
\qquad
\subfloat[$T=200$]{
  \centering
  \includegraphics[width=0.25\linewidth]{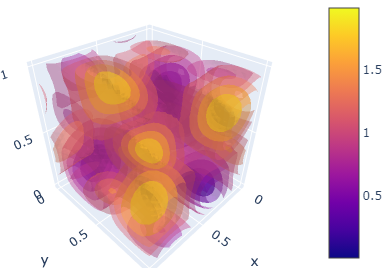}
}
\qquad
\subfloat[$T=500$]{
  \centering
  \includegraphics[width=0.25\linewidth]{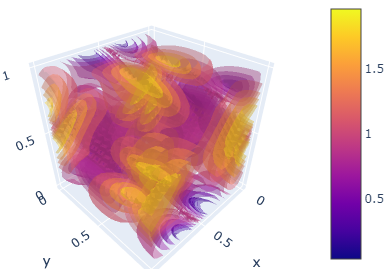}
}
\caption{Evolution of the $3D$ sinusoidal initial condition defined in Eq.\ \eqref{eq:sin3d} and 3D swirl velocity field defined in Eq.\ \eqref{eq:swirl3d} with CUDA-Q simulator.}\label{fig:sim_results}
\end{figure*}

Once we retrieve the final state, we may measure some observable that is of interest to us. The overall probability of success is only dependent on the $L_2$ norm of the initial and final density functions. Note that the changes made to the algorithm discussed in Sec.~\ref{sec:improve_prob_success} are critical for this method to scale efficiently. Applying multiple timesteps in a single circuit without these changes would result in an exponentially decaying overall probability of success given by
\begin{equation}
    \text{Pr}(|0\rangle_D)=\frac{\|\Phi_{T}\|^2}{2^{2Tn_d}\|\Phi_0\|^2}.
\end{equation}
As shown in Fig.~\ref{fig:succ_prob}, this $T$ dependence requires an exponential number of measurement shots while our improved method always results in a constant success probability at long T.

Putting together the results of the previous sections, we now see that the total circuit complexity of our algorithm scales polynomially time and poly-logarithmically in grid size. Overall, we have $\texttt{time} \sim \mathcal{O}\big(M\log(N) T\frac{||\Phi_T||^2}{||\Phi_0||^2 }+ \log(N)\chi^2 \big) $, where $\chi$ is the bond dimension of the input MPS state which we assume scales at worst as $\text{polylog}(N)$, and $\texttt{space} \sim \mathcal{O}\big(M+2D\log_2(L)\big)$.

\section{Generalizing to non-uniform velocity fields}

Ref. \cite{altair} demonstrates how arbitrary velocities can be implemented with the QLBM approach of \cite{Budinski2021}. This task becomes a bit more involved when we use the LCU approach in Section \ref{sec:improve_prob_success}.

To incorporate position dependent velocities, we need to make changes to the PREP and PREP$^\dagger$ operators.

The new PREP operator will be constructed such that

\begin{equation}\label{eq:prep_define}
    \text{PREP}|x\rangle_G|0\rangle_D=|x\rangle_G\sum_i\sqrt{k_i(x)}|i\rangle,
\end{equation}
and PREP$^\dagger$ will be renamed UNPREP (it is no longer just the adjoint of PREP) and will be constructed such that
\begin{equation}\label{eq:unprep_define}
    \text{UNPREP}^\dagger|x\rangle_G|0\rangle_D=|x\rangle_G\sum_i\sqrt{k_i(x-c_i\Delta t)}|i\rangle.
\end{equation}

Here we note that this operator is only unitary if
\begin{equation}
\sum_i |k_i(x-c_i\Delta t)|=1
\end{equation}
for all values of $x$. This algorithm is summarized in Alg. \ref{alg:qlbm_nonuni}.

The circuit implementation of these PREP and UNPREP operators depend on the specific velocity field. We now present an illustration for a simple non-uniform velocity field and its circuit implementation.

\begin{figure*}
\centering
\subfloat[$T=0$]{
  \centering
  \includegraphics[width=0.19\linewidth]{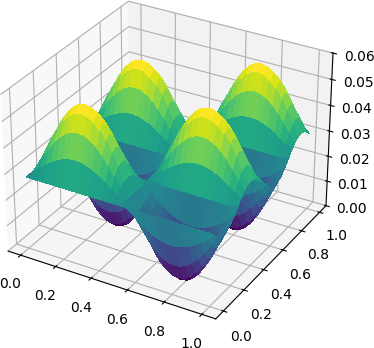}
}
\qquad
\subfloat[$T=15$]{
  \centering
  \includegraphics[width=0.19\linewidth]{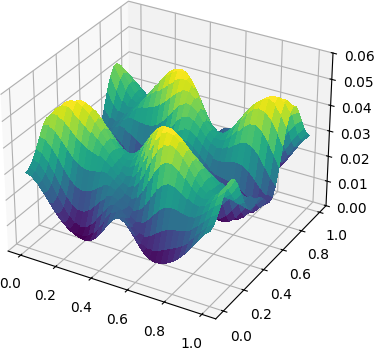}
}
\qquad
\subfloat[$T=30$]{
  \centering
  \includegraphics[width=0.19\linewidth]{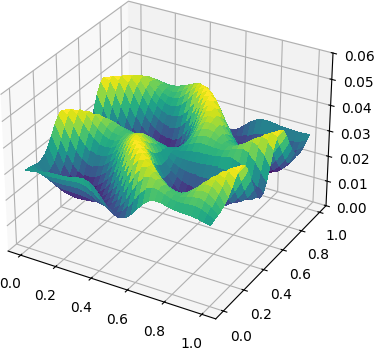}
}
\caption{
Evolution of the $2D$ sinusoidal initial condition defined in Eq.\ \eqref{eq:sin2d} and 2D swirl velocity field defined in Eq.\ \eqref{eq:swirl2d} with CUDA-Q simulator.}
\label{fig:sim_results2}
\end{figure*}

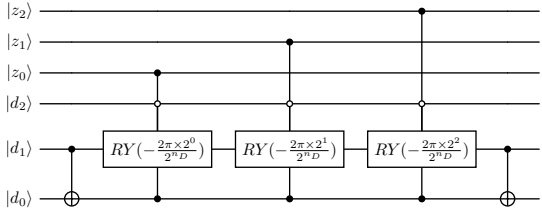
\begin{figure}
    \centering
    \resizebox{0.40\textwidth}{!}{
    \begin{quantikz}
        \lstick{$|z_2\rangle$}&&&&\ctrl{4}&&\\
        \lstick{$|z_1\rangle$}&&&\ctrl{3}&&&\\
        \lstick{$|z_0\rangle$}&&\ctrl{2}&&&&\\
        \lstick{$|d_2\rangle$}&&\octrl{1}&\octrl{1}&\octrl{1}&&\\
        \lstick{$|d_1\rangle$}&\ctrl{1}&\gate{RY(-\frac{2\pi\times2^0}{2^{n_D}})}&\gate{RY(-\frac{2\pi\times2^1}{2^{n_D}})}&\gate{RY(-\frac{2\pi\times2^2}{2^{n_D}})}&\ctrl{1}&\\
        \lstick{$|d_0\rangle$}&\targ{}&\ctrl{-1}&\ctrl{-1}&\ctrl{-1}&\targ{}&\\
    \end{quantikz}}
    \caption{Circuit implementation of a $RBS(2\pi z)$ gate controlled on the $|0\rangle$ state of $d_2$, whose angle depends on the state of qubits $\ket{z_0z_1z_2}$.}
    \label{fig:r_circ}
\end{figure}

\begin{algorithm}
\caption{QLBM generalized for non-uniform flows}\label{alg:qlbm_nonuni}
\begin{algorithmic}[1]
\Require Initial state $|\Phi\rangle_G|0\rangle_D$
\Ensure Final state $|\Phi_{t+1}\rangle$

\State Apply PREP as defined in Eq. \eqref{eq:prep_define}.
\begin{equation*}
    \longrightarrow\frac{1}{\|\Phi_t\|}\sum_x\sum_i\sqrt{k_i(x)}\Phi(x,t)|x\rangle_G|i\rangle_D
\end{equation*}
\State Apply the streaming operator $U_S$ from Eq. \eqref{eq:stream_defined}.
    \begin{align*}
        \longrightarrow&\frac{1}{\|\Phi_t\|}\sum_x\sum_i\sqrt{k_i(x)}\Phi(x,t)|x+c_i\Delta t\rangle_G|i\rangle_D\\
        =&\frac{1}{\|\Phi_t\|}\sum_{x,i}\sqrt{k_i(x-c_i\Delta t)}\Phi(x-c_i\Delta t,t)|x\rangle_G|i\rangle_D
    \end{align*}
\State Apply the UNPREP operator as defined in Eq. \eqref{eq:unprep_define}.
    \begin{equation*}
        \longrightarrow\frac{\|\Phi_{t+1}\|}{\|\Phi_t\|}|\Phi_{t+1}\rangle_G|0\rangle_D+|\chi\rangle
    \end{equation*}
    where $(\mathbbm{1}_G\otimes\langle0|_D)|\chi\rangle=0$.
\State Measure the direction qubits and post-select for the $|0\rangle$ state
\begin{equation*}
    \longrightarrow|\Phi_{t+1}\rangle_G
\end{equation*}
\end{algorithmic}
\end{algorithm}

\subsection*{Illustration}

Consider a D3Q7 model with the following velocity field:


\begin{equation}\label{eq:swirl3d}
    (u_x,u_y,u_z)=\left(\frac{1}{3}\sin(-2\pi z),\frac{1}{3},\frac{1}{3}\sin(2\pi x)\right),
\end{equation}

for $x$, $y$, $z \in \left[0,1\right)$.

The PREP operator should be constructed such that,
\begin{align}
     &\text{PREP}|(x,y,z)\rangle_G|0\rangle_D\nonumber \\
     &=|(x,y,z)\rangle_G\sum_i\sqrt{k_i(x,y,z)}|i\rangle_D  \\
     & =|(x,y,z)\rangle_G\otimes \begin{bmatrix}
         1/2\\
         \frac{1}{2}\cos(\frac{\pi}{4}+\pi z)\\
         \frac{1}{2}\sin(\frac{\pi}{4}+\pi z)\\
         1/2\\
         0\\
         \frac{1}{2}\cos(\frac{\pi}{4}-\pi x)\\
         \frac{1}{2}\sin(\frac{\pi}{4}-\pi x)\\
         0
     \end{bmatrix}.\label{eq:prep_final}
\end{align}
Note that here we have used a dense encoding for the PREP operator. In practice however, when using real hardware, it is often a better choice to use a one-hot encoding to decrease the overall depth after transpilation at the expense of marginally more qubits (2 extra qubits in this example).

Here, we evaluate $\{k_i\}_i$ using Eq.\ \eqref{eq:k_define} with the following ordering of $\{c_i\}_i$:
\begin{align}
    [&(0,0,0),(1,0,0),(-1,0,0),(0,1,0),\nonumber\\
    &(0,-1,0),(0,0,1),(0,0,-1)] .
\end{align}
In the case of this velocity field, UNPREP equals PREP$^\dagger$.

Now we discuss the implementation of this operator.
\begin{itemize}
    \item Let the grid qubits be in a computational basis state $|(x,y,z)\rangle=|x\rangle|y\rangle|z\rangle$ and the direction qubits be in the $|0\rangle$ state.
    \item We use standard state preparation techniques to create the following state $|(x,y,z)\rangle_G|k_0\rangle_D$.

    where $|k_0\rangle$ has the following explicit form:
    \begin{align}
    |k_0\rangle=\begin{bmatrix}
        \frac{1}{2} &
        \frac{1}{2\sqrt{2}} &
        \frac{1}{2\sqrt{2}} &
        \frac{1}{2} &
        0 &
        \frac{1}{2\sqrt{2}} &
        \frac{1}{2\sqrt{2}} &
        0
    \end{bmatrix}
    \end{align}

    \item As shown in the circuit depicted in Fig. \ref{fig:r_circ}, we can apply $RBS(2\pi z)$ as defined in \cite{rbs} on the two least significant direction qubits controlled on the $|0\rangle$ state of the most significant direction qubit.

    \item We can similarly apply $RBS(2\pi x)$ controlled on the $|1\rangle$ state of the most significant direction qubit to get our final desired state as seen in Eq. \eqref{eq:prep_final}.
    
\end{itemize}

\subsection{Simulation results}
We validated the proposed QLBM algorithm for the following 3D initial condition on a $512\times512\times512$ grid and $500$ timesteps,

\begin{equation}\label{eq:sin3d}
    |\Phi_0\rangle=\sin(2\pi x)\sin(2\pi y)\sin(2\pi z)+1,
\end{equation}
where $x,y,z\in[0,1)$. This 32-qubit simulation used $8$ Nvidia A100 GPUs, leveraging the CUDA-Q library \cite{cudaq} for efficient GPU utilization. The results are shown in Fig. \ref{fig:sim_results}.
\par We also show results in Fig. \ref{fig:sim_results2}, for a similar $32\times32$ D2Q5 example with the velocity field,
\begin{equation}\label{eq:swirl2d}
    (u_x,u_y)=\left(\frac{1}{3}\sin{(-2\pi y)},\frac{1}{3}\sin{(2\pi x)}\right),
\end{equation}
and initial state,
\begin{equation}\label{eq:sin2d}
    |\Phi_0\rangle=\sin(2\pi x)\sin(4\pi y)+1.
\end{equation}

We verify without showing here in interest of brevity, that the results presented tally exactly with the classical LBM solution. These results capture the expected behaviour of a swirling motion with diffusion, and serve to demonstrate generalization of the proposed algorithmic techniques to $3D$ data, and non-uniform velocity fields.

\section{Conclusion}
In this work, we introduced a series of algorithmic advancements that enable state-of-the-art QLBM implementations on simulators to be executed on current-generation quantum hardware. Leveraging these techniques, we presented the first demonstration of the two-dimensional advection-diffusion equation on a quantum processor, showing excellent agreement with classical benchmarks. Our approach addresses prohibitive limitations of prior work, including challenges in state preparation and observable readout, as well as the substantial circuit depth of collision and streaming operations. Through circuit-level optimizations and error mitigation strategies, we achieved significant reductions in the depth and noise sensitivity, yielding high-fidelity results on NISQ hardware.
\par
We further extended our methodology to three-dimensional domains and non-uniform velocity fields, validating these developments through large-scale numerical simulations involving up to 32 qubits,  thereby laying the groundwork for applying QLBM to a broader class of linear scalar transport problems. \par
Future work will focus on scaling quantum hardware implementations to support longer time evolutions and higher-dimensional systems. When extended to overcome challenges of non-linearity and complex boundary conditions, our approach provides a foundation for developing industrial-scale quantum-enabled computational physics solvers.

\section*{Acknowledgement}
The authors would like to thank Masako Yamada for facilitating the logistics of this collaborative project.

\bibliographystyle{unsrt}
\bibliography{references, QLBM_IEEE_refs}

\end{document}